# Optimal Geometric Design of Thermoelectric Metamaterials for Enhancing Power Generation: An Interpretative Approach

Xanthippi Zianni

Dept. of Aerospace Science and Technology, National and Kapodistrian University of Athens, Greece.

**ABSTRACT**

Thermoelectric metamaterials featuring width modulation through constrictions (*constricted* geometries) have emerged as a promising approach for improving heat management and thermoelectric performance. Through a combination of theoretical calculations, analytical formalism, and validation against experimental data, it is shown that thermoelectric performance in such geometries is governed by two fundamental mechanisms of pure geometrical origin: (i) a characteristic scaling behavior of resistance with *Transmissivity*, and (ii) the critical formation of the *Constriction Thermal Resistance*. Hourglass-shaped thermoelectric legs—identified as optimal in recent experiments—are found to exhibit the same underlying transport mechanisms observed in other *constricted* profiles, including single and multiple sharp constrictions. The commonly used *Geometric Parameter* is found to be insufficient for capturing the full influence of geometry on transport, whereas *Transmissivity* serves as a robust descriptor of *constricted* geometry, independent of material choice or device operating conditions. A universal scaling formalism is derived linking electrical and thermal resistances, along with key thermoelectric performance metrics, to the *Transmissivity*. A unified optimization framework is also developed for composite legs, incorporating both *constricted* material and contact electrodes. This framework indicates that previously reported performance gains may be largely attributed to contact resistance, rather than geometry alone. *Transmissivity* is established as a key geometric descriptor, enabling generalized design principles and global optimization criteria for enhancing thermoelectric power generation. This analysis elucidates new avenues in the design of thermoelectric metamaterials for efficient energy conversion.

## I. INTRODUCTION

Thermoelectric (TE) energy conversion transforms waste heat into electrical power, addressing energy inefficiency and environmental concerns[1-6]. As energy demand rises—driven by microelectronics, AI, and quantum tech—thermoelectrics provide compact, solid-state solutions for decentralized power[7-8]. They harvest low-grade heat (e.g., exhaust, body heat), enabling battery-free operation of sensors and processors in autonomous systems[9-10]. By reclaiming thermal energy, TE devices support green strategies, reduce thermal pollution, and aid decarbonization[11-12]. Their scalability, durability, and silent function make them ideal for next-gen, sustainable technologies.

TE performance depends on material traits—mainly the figure of merit ($ZT$), tied to Seebeck coefficient $S$, electrical conductivity $\sigma$, and thermal conductivity $k$ as $ZT = \sigma S^2 T/k$ — and device-level aspects like geometry and contact resistance[1,3]. While high-$ZT$ materials exist[13-21], integration into efficient modules faces hurdles like interface losses, limited scalability, and thermal/electrical mismatches[22-25].

Enhancing TE performance requires optimizing the underlying transport properties-specifically, reducing $R_{el}$, increasing $S$, and raising the thermal resistance $R_{th}$. A higher $R_{th}$ is essential for maintaining a robust temperature gradient across TE legs. Two primary material-level strategies have been pursued: (i) the discovery of new materials with superior properties and higher figure of merit $ZT$, and (ii) the structural engineering of materials at the nano- and microscale to enhance their transport properties, thereby surpassing the performance of their bulk counterparts. In typical cuboid TE leg design, the optimization of $R_{el}$ and $R_{th}$ is carried out by adjusting the *aspect ratio* of the leg, owing to the inherent trade-off between the leg's length and its cross-sectional area[1,3,26].

An emerging strategy in TE research is geometry modulation. This concept proposes tailoring the physical geometry of the material of the TE leg to controlling electrical and thermal transport beyond what is achievable through material composition or structural processing alone. Non-cuboid structures with variable cross-sections—especially *pyramidal* geometries—have attracted interest in TE research since its early stages[27-29]. However, these configurations remained relatively underexplored until recent advancements in metamaterials reignited attention. In particular, width-modulated metamaterials featuring alternating constrictions and expansions (*constricted* geometries) have been proposed as a means to enhance TE efficiency by geometrically tuning electrical and thermal transport properties[30-34]. A resurgence of interest has sparked, leading to renewed theoretical investigations into both *pyramidal*[35-38] and *constricted* geometries[39-41], which have demonstrated the potential for significantly improved TE performance. This strategy has, though, also not been validated in device modules, where complex geometry fabrication limitations, integration complexity, and interface quality remain key obstacles[42-45]. Recent advances in 3D microfabrication and additive manufacturing have begun to make the experimental realization viable[46-48]. These findings have spurred a growing body of research focused on TE leg configurations with non-cuboid geometries[49-53]. Future progress in this area will rely on multi-parameter optimization of TE devices incorporating such non-standard leg shapes[49-51]. A critical step in this process is the geometric optimization of metamaterials with respect to key TE performance metrics. Numerical studies have investigated a variety of non-cuboid designs (Fig. 1), demonstrating that variable cross-sections can significantly improve performance across a range of boundary conditions and design constraints, such as fixed volume or surface area[39-41,48-53]. A recent study implemented a design strategy to determine the optimal geometry of TE legs for high-temperature power generation using $Cu_2Se$, fabricated via an extrusion-based 3D printing process[48]. Eight leg geometries—including cuboid, truncated pyramid, Y-shaped, and multi-hollow rectangular designs—were evaluated under diverse operating conditions (Fig.1b). Among them, the hourglass-geometry demonstrated significantly higher power output and efficiency compared to the conventional cuboid geometry, confirming previous theoretical predictions[31,32] on *constricted* geometries (Figs.1c and 1d).

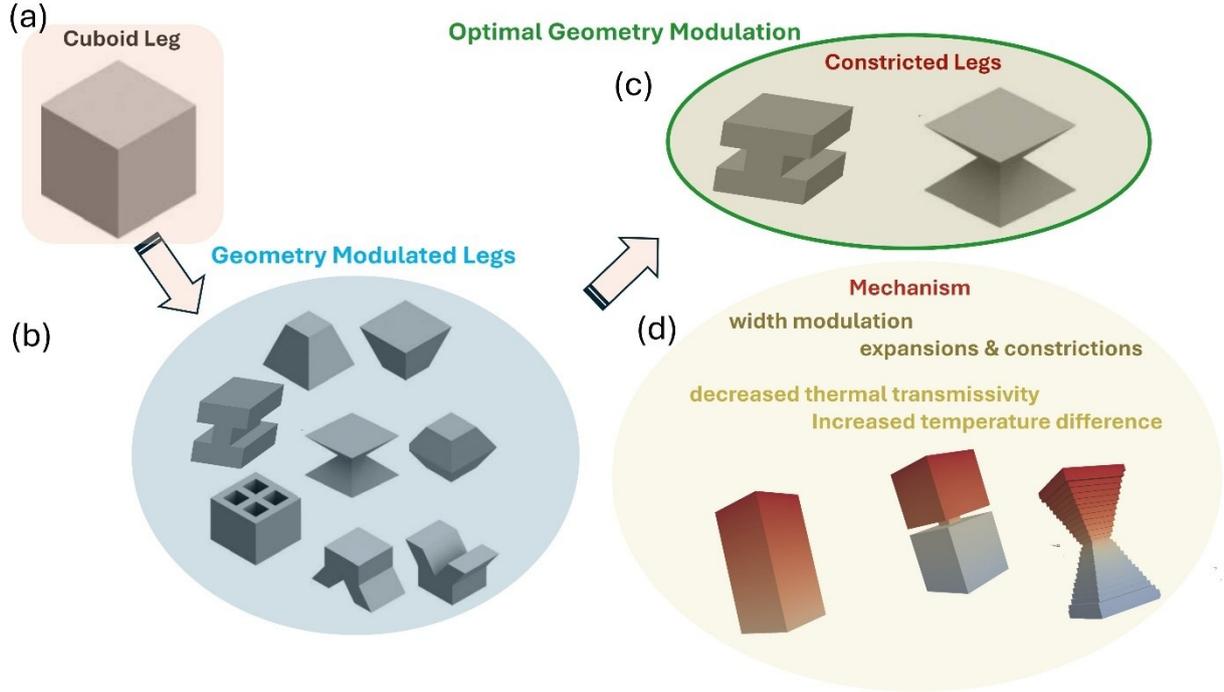

**FIG.1. Geometric design, optimization and physics mechanism**. (a) The cuboid TE leg geometry, (b) Fabricated geometry-modulated non-cuboid TE legs, (c) Constricted legs, as the optimal geometry design, (d) Underlying physics mechanism.

*Constricted* geometry modulation affects transport processes on two distinct levels: material and structural. At the material level, it influences thermoelectric and thermal transport by altering the energy states of electrons and phonons, as well as their scattering behavior, primarily due to quantum confinement effects[30,54-56]. At the structural level, transport is suppressed through reduced *Transmissivity*—a purely geometrical effect that is fundamentally different from traditional boundary or interface scattering mechanisms[31-32]. For *constricted* geometries, *Transmissivity* is expressed[32] as the ratio of the constriction cross-sectional area $A_C$ to that of the expansion $A$ (Fig.2a):

$$Tr = \frac{A_C}{A} \qquad (1)$$

In this work, the symbol *Tr* will be used to denote this geometric ratio, rather than to represent the more general concept of *Transmissivity*. The above expression must be appropriately adapted to reflect the specific features of the specific geometry modulation design.

The concept of *Transmissivity*—entirely governed by the geometry-modulation profile—was originally introduced as an intuitive framework for studying nanoscale thermal transport in *constricted* metamaterials[31]. Subsequent studies employing rigorous phonon Monte Carlo simulations validated these initial insights and uncovered distinct features associated with this mechanism[32]. A key outcome was the identification of a characteristic scaling relationship between thermal conduction and *Transmissivity*[31-32], demonstrating it as an intrinsic transport property of *constricted* geometries— rather than a simple geometric ratio that might also describe other shapes (e.g., *pyramidal*) without capturing the same underlying physical mechanism.

Geometry optimization for maximum TE performance hinges on a deep physical understanding of geometry-modulated structures—a nontrivial challenge. It demands the identification of quantitative links between geometric metrics and physical transport properties. Although enhanced TE performances were attributed to enlarged Geometric Parameter (*GP*) [48-51], such enhancements lacked general applicability across different geometries due to the absence of universal quantitative relationships. The present study addresses this need by demonstrating a *universal scaling* relationship of electrical and thermal transport, as well as key TE performance metrics, with *Transmissivity* in *constricted* geometries – though calculations validated with experimental data and analytical formalism.

Calculations on *constricted* TE legs with hourglass shape—identified as optimal in the experimental[48] study—demonstrate that transport and TE performance are governed by the same mechanisms observed in other *constricted* geometries, including single and multiple sharp constriction modulation profiles[31-32]: (i) the characteristic *scaling behavior* of thermal resistance with *Transmissivity*, and (ii) the critical formation of the *Constriction Thermal Resistance (CTR)*. Experimental[48] observations on hourglass TE legs are interpreted in terms of these mechanisms.

The theoretical model is described in Section II. Geometry modulation as a design strategy is investigated in Section III.A, demonstrating that *Transmissivity*, rather than the *GP*, is the appropriate metric to capture the effect of geometry modulation in *constricted* TE legs. The relationship between *Transmissivity* and performance is examined in Section III.B, showing a universal scaling behavior of electrical and thermal resistances, as well as TE performance metrics, with *Transmissivity* – supported by calculations and analytical formalism. In Section III.C, a universal scaling formalism is developed for TE transport and performance with *Transmissivity*. Section III.D presents a unified optimization framework to explain output power in real TE devices comprising the *constricted* material and contact electrodes. Finally, in Section III.E, *constricted* geometry optimization is interpreted in terms of the *Constriction Therma Resistance CTR*.

## II. THEORETICAL MODEL

Calculations were carried out using the finite element method in which the structure is discretized into individual cells. An electrical potential difference $\Delta V$ and a temperature difference $\Delta T$ are imposed on the two opposite sides of the structure. The electrical current densities $\boldsymbol{J}$ and heat flux $\boldsymbol{q}$ are calculated using the following definitions[1,3,57]:

$$\boldsymbol{J} = -\sigma(\nabla V + S\nabla T) \qquad (2a)$$

$$\boldsymbol{q} = -k\nabla T + ST\boldsymbol{J} \qquad (2b)$$

The profiles of the electrical potential $V$ and the absolute temperature $T$ are determined by applying electric current continuity and energy conservation conditions[57]:

$$\nabla \cdot \boldsymbol{J} = 0 \tag{3a}$$

$$\nabla \cdot \boldsymbol{q} = -\nabla V \cdot \boldsymbol{J} \tag{3b}$$

The simulation method was validated by comparison with recent experimental data[48] on Cu$_2$Se cuboid and non-cuboid TE legs under fixed temperature and convective heat flux conditions, as well as with analytical formalism. The calculations used experimental transport properties $\sigma$, $S$ and $k$, fully accounting for their temperature dependence. Consequently, the effects of the positional variation of $\sigma$, $S$ and $k$, arising from the calculated spatial temperature distribution within the leg (including the Thomson effect) are incorporated. Excellent agreement was observed between simulations, measurements, and the analytical models, as detailed in Section III. Systematic calculations were performed by varying the cross-sectional areas $A$ and $A_c$ and the convection coefficient $h$; representative results are presented.

TE devices are based on composite legs — consisting of the contact electrodes and the material slab (bare leg). In Sections III.A, III.B and III.E, the calculations pertain to bare legs and are referred to simply as 'legs'. Section III.D, which is dedicated to comparison with experimental data, presents calculations for both composite and bare legs.

## III. RESULTS AND DISCUSSION

**A. Geometry modulation as design strategy of TE legs**

The performance of a TE leg is typically evaluated using the TE efficiency $\eta$ and the maximum output power $P_{max}$:

$$\eta = \frac{P_{el}}{Q_{in}} \tag{4}$$

$$\eta_{max} = \frac{\Delta T}{T_h} \frac{\sqrt{1+ZT}-1}{\sqrt{1+ZT}+\frac{T_c}{T_h}} \tag{4a}$$

$$P_{max} = \frac{V_{OC}^2}{4\,R_{el}} = \frac{S^2 \Delta T^2}{4\,R_{el}} \tag{5}$$

where $P_{el}$ and $Q_{in}$ are respectively the electrical output power and the incoming thermal power. $n_{mas}$ is the maximum efficiency. $T_h$ and $T_c$ denote the temperature at the hot side and the cold side of the leg respectively. $P_{max}$ is the maximum electrical output power, $V_{OC}$ is the open circuit voltage, $\Delta T$ is the temperature difference across the sides of the leg and $R_{el}$ is its electrical resistance.

To maximize TE performance, it is essential to optimize both $R_{el}$ and $R_{th}$. $R_{el}$ must be small for high output power [Eq. (5)]. $R_{th}$ must be large to maintain a large $\Delta T$ and maximize efficiency and output power [Eqs. (4a) and (5)]. Resistances are influenced by the material's electrical conductivity $\sigma$ and thermal conductivity $k$, as well as its geometry. In the case of a *cuboid* material:

$$R_o^{el} = \frac{1}{\sigma}\frac{L}{A} \tag{6a}$$

$$R_o^{th} = \frac{1}{k}\frac{L}{A} \tag{6b}$$

where $R_o^{el(th)}$ denotes the electrical (thermal) resistance of the cuboid material. $L$ is the length of the leg and $A$ is its cross-sectional area. Therefore, in addition to optimizing $\sigma$ and $k$, the dimensions $L$ and $A$ serve as standard design variables in conventional TE leg optimization.

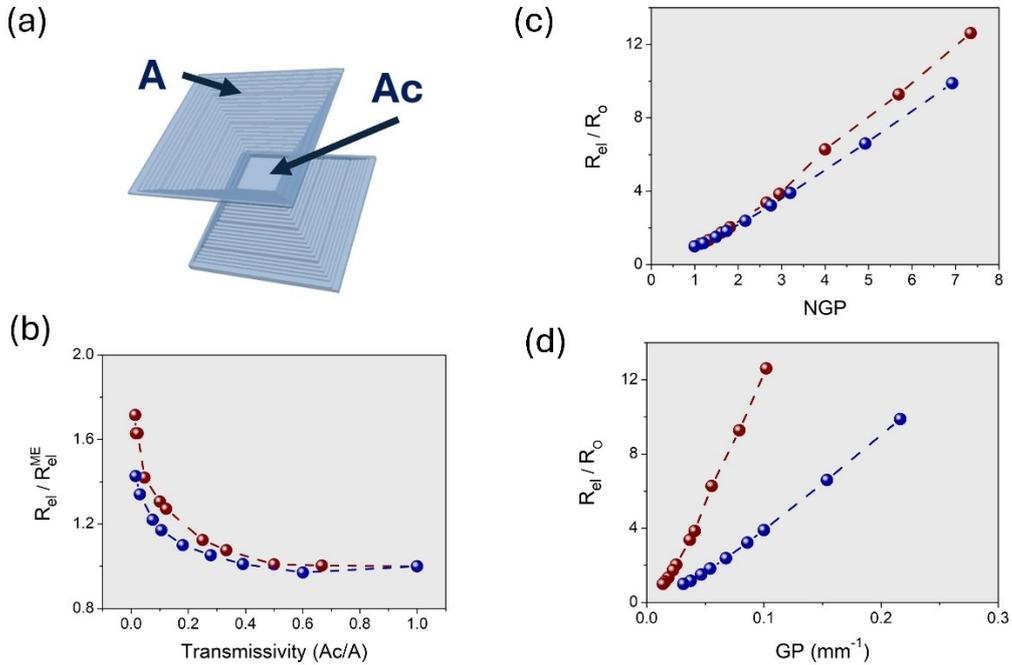

**FIG.2 Geometry modulation metrics.** (a) Schematic of the hourglass geometry showing the outer cross-sectional area $A$ and the constriction cross-sectional area $A_C$. (b) Ratio of the calculated electrical resistance $R_{el}$ and estimated electrical resistance $R_{el}^{ME}$ versus *Transmissivity*. (c) Ratio of the hourglass resistance $R_{el}$ to that of the corresponding cuboid $R_o$ plotted against the Normalized Geometric Parameter (NGP). (d) Same resistance ratio plotted against the Geometric Parameter (GP).

However, in geometry-modulated materials, *L* and *A* no longer capture the complexity of the actual shape and are thus not meaningful optimization parameters. To address this, Choo et al.[48] used the modified expressions (ME) for non-cuboid materials:

$$R_{el}^{ME} = \int_0^L \frac{dl}{\sigma A(l)} \tag{7a}$$

$$R_{th}^{ME} = \int_0^L \frac{dl}{k A(l)} \tag{7b}$$

and adopted the *Geometric Parameter* (*GP*) as a more suitable criterion to for maximizing $P_{max}$ than the aspect ratio of a cuboid leg[48]:

$$GP \equiv \int_0^L \frac{dl}{A(l)} \tag{8}$$

Nonetheless, calculations show that the modified expressions do not hold for *constricted* geometries:

$$R_{el} \neq \int_0^L \frac{dl}{\sigma A(l)} \tag{9a}$$

$$R_{th} \neq \int_0^L \frac{dl}{k A(l)} \tag{9b}$$

where $R_{el(th)}$ denotes the calculated electrical (thermal) resistance of the leg.

If Eqs. (7) were valid, the ratio $R_{el(th)}/R_{el(th)}^{ME}$ would equal to 1. However, the plots of $R_{el}/R_{el}^{ME}$ versus the *Transmissivty* of *constricted* legs in Fig.2b clearly show that this is not the case: the ratio deviates from 1, and the deviation increases as *Transmissivity* decreases. This indicates that the violation of Eqs. (7) becomes more pronounced as the *constricted* geometry departs further from the cuboid shape.

The failure of Eqs (7) stems from the assumption that the resistance of a variable cross-section material can be obtained by integrating along its length as if it were composed of a series of constant cross-section slices. This implicitly assumes that current remains parallel to the resistor's axis under the applied field—hence not parallel to the resistor's sides—which would require current to pass through the sides, contradicting the current continuity condition (Eq.(3a)). In contrast, this boundary condition is correctly satisfied in the finite element calculations. The failure of this in-series resistor model for variable cross-section resistors has been noted previously for conical[58] and *constricted* geometries[31].

Further insight can be gained in terms of the *Geometric Parameter Deviation Ratio* (*GPDR*) and the *Normalized Geometric Parameter* (*NGP*), defined as:

$$GPDR \equiv \left(\frac{R}{R_o}\right)/NGP \tag{10}$$

$$NGP \equiv \frac{A}{L}\int_0^L \frac{dl}{A(l)} = \frac{A}{L}GP \tag{11}$$

where the subscripts *el* and *th* have been omitted for simplicity. *R* stands for the resistance of the *constricted* material and $R_o$ for the resistance of the corresponding cuboid material.

For uniform thermal and electrical conductivities *k* and *σ*, Eqs. (7) yield:

$$R^{ME} = R_o \frac{A}{L}GP = R_o\, NGP \tag{12}$$

The combination of Eqs. (10) and (12) gives:

$$GPDR = \frac{R}{R^{ME}} \tag{13}$$

Eq. (13) shows that *GPDR* serves as metric of the violation of Eqs. (7), as its deviation from unity increases in proportion to the extent of the violation, similarly to the ratio $R/R^{ME}$. *GPDR* is greater than 1 when the resistance of the *constricted* geometry resistors exceeds that predicted by Eqs. (7) (Fig.2b).

Eq. (12) shows that $R^{ME}$ scales with *NGP*. Thus, if Eqs. (7) were valid, *R* would match $R^{ME}$ and exhibit the same scaling, implying that *NGP* alone governs the effect of *constricted* geometry and could serve as an optimization parameter. However, Fig.2c shows that $R_{el}/R_o$ does not scale with *NGP* further confirming the breakdown of Eqs. (7). In Fig.2d, $R_{el}/R_o$ is plotted against *GP*, showing no scaling with *GP* either. In contrast, $R_{el}/R_o$ scales consistently with *Transmissivity*, following the same scaling relationship for electrical and thermal resistances (Fig.3a). This invariance to transport carrier type (electron/phonon) and intrinsic material transport properties shows that the change in resistance *R* due to *constricted* geometry - relative to $R_o$ of the uniform structure - is a purely geometric effect governed entirely by the *Transmissivity*, Consequently, *Transmissivity* is appropriate geometric descriptor for optimizing the TE performance of *constricted* TE legs.

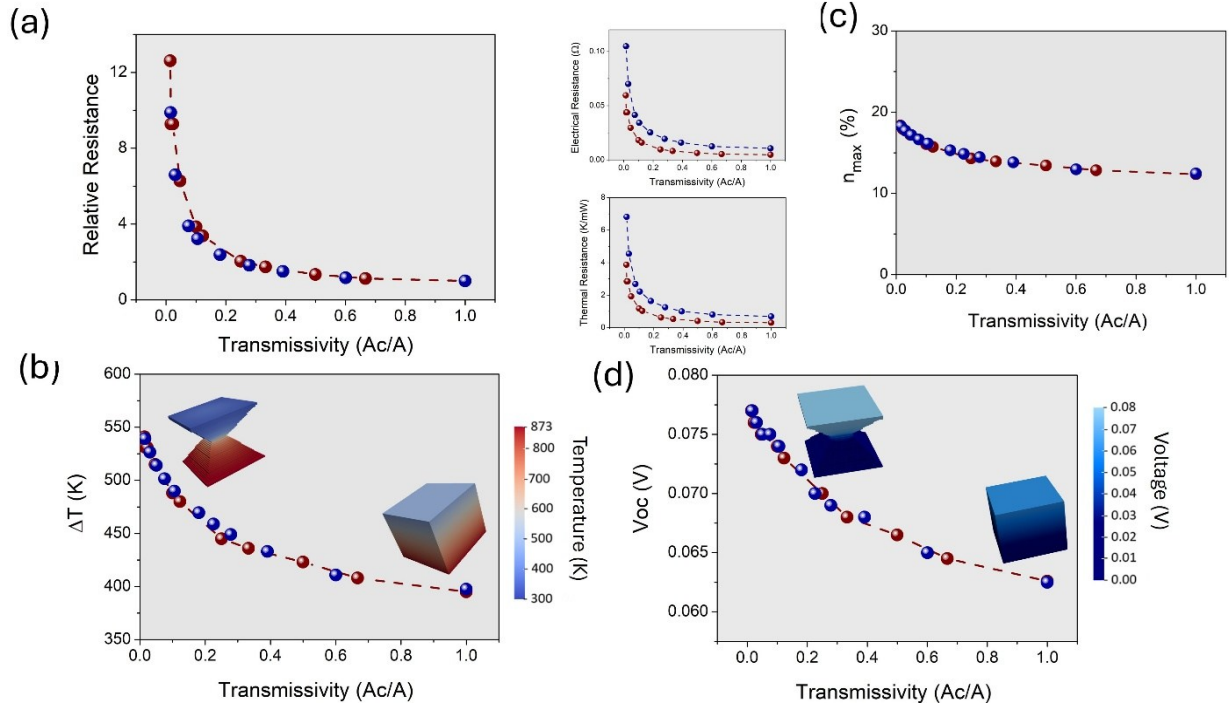

**FIG. 3. Scaling Behavior with Transmissivity.** (a) Resistance relative to that of the cuboid resistance (R/Ro) and absolute resistance values for electrical and thermal transport. (b) Temperature difference $\Delta T$ and characteristic temperature profiles. (c) Maximum efficiency $n_{max}$. (d) Open-circuit voltage $V_{OC}$ and characteristic output voltage profiles. Calculations are for two sets of $Cu_2Se$ hourglass legs with $L$= 4mm: one with cross-sectional area $A$=4X4 mm$^2$ (blue symbols) and the other with $A$=6X6 mm$^2$ (red symbols), each with varying *Transmissivities*.

### B. Transmissivity and performance of constricted TE legs

The efficiency of a material with $\sigma$, $S$ and $k$, is determined by the temperature difference $\Delta T$ across the sides of the leg. $\Delta T$ reaches it's maximum when the temperatures at both sides are fixed. However, under practical operating conditions, the boundary temperatures are often not constant, as they are influenced by convective heat exchange. In such cases, the temperature at a contact is governed by the material's thermal resistance and the strength of convective heat flow, typically characterized by the convection heat transfer coefficient, $h$. A higher value of $h$ indicates more efficient heat exchange, driving the contact temperature closer to the ambient value. For $h \to \infty$, the contact temperature equals that of the ambient environment. Conversely, as $h$ decreases, the achievable temperature difference across the leg diminishes—posing a well-known limitation in TE applications with weak convective flow.

Under convective operating conditions, the temperature difference $\Delta T$ across *constricted* legs exceeds that of *cuboid* legs due to the smaller *Transmissivity* and consequent reduced thermal conduction[31,32]. A similar increase was measured in the hourglass geometry and was attributed to larger *GP*s influencing $R_{th}$ and the effective heating/cooling surface areas[48]. However, present analysis does not support this interpretation. Calculations reveal that, for fixed *h* and intrinsic material properties, *ΔT* is entirely governed by the *Transmissivity*. Specifically, *ΔT* increases as *Transmissivity* decreases—reflecting the increased thermal resistance (Fig.3b). Most importantly, *ΔT* scales directly with the *Transmissivity* (Fig.2c). This implies that *constricted* legs with different surface areas can exhibit identical temperature gradients, provided they have the same *Transmissivity*.

The calculated scaling dependence of *ΔT* on *Transmissivity* is validated by the following analytical formalism. Consider first a cuboid material of cross-sectional area *A* and length *L*, where *Tr*=1. For a hot-side temperature $T_h$, the cold-side temperature $T_c$ under convective conditions is determined by the thermal conductivity *k*, the length *L*, and the convection coefficient *h*, according to the standard steady-state heat transfer solution:

$$T_c = T_h - \frac{hL}{k + hL}(T_h - T_a) \tag{14}$$

where $T_a$ is the ambient temperature.

This expression explicitly shows that *ΔT* is independent of *A* for cuboid geometries (*Tr*=1), consistent with calculations in Fig.3b.

For non-cuboid geometries, Eq.(14) can be extended as follows:

$$T_c = T_h - \frac{hL}{k_{eff} + hL}(T_h - T_a) \tag{15}$$

in terms of the effective thermal conductivity $k_{eff}$, defined from the non-cuboid material's thermal conductance $G_{th}$ through the relation:

$$G_{th} = k_{eff} \frac{A}{L} \tag{16}$$

The intrinsic thermal conductivity *k* and the thermal conductance of the cuboid material $G_o^{th}$ are related by:

$$G_o^{th} = k \frac{A}{L} \tag{17}$$

Equations (16) and (17) give:

$$\frac{k_{eff}}{k} = \frac{G_{th}}{G_o^{th}} = \frac{R_o^{th}}{R_{th}} \tag{18}$$

Then, from Equations (15) and (18), it is obtained:

$$\Delta T = \frac{hL/k}{k_{eff}/k + hL/k}(T_h - T_a) \tag{19}$$

or equivalently expressed, in terms of $\Delta T_{max} \equiv T_h - T_a$:

$$\Delta T = \frac{hL/k}{R_o^{th}/R_{th} + hL/k}\Delta T_{max} \tag{20}$$

Eq. (20) applies to both cuboid and non-cuboid materials. For *constricted* geometries, it shows that $\Delta T$ is independent of the cross-sectional area $A$, and - under convective conditions- scales with *Transmissivity* according to the same relation as $R_o^{th}/R_{th}$ (Figs.3a and 3b). This scaling arises purely from geometry modulation, confirming our calculations and interpretation, and explaining why the variation in thermal resistance with $A$ shown in Fig. 3b does not produce a corresponding change in $\Delta T$ when *Transmissivity* is fixed.

Eq. (20) exactly matches the calculations in Fig.3b for the experimental parameters $h$=200 Wm$^{-2}$K$^{-1}$, $L$=4 mm, $k$=0.4 Wm$^{-1}$K$^{-1}$, and the calculated $R_o^{th}/R_{th}$ shown in Fig.3a. (The corresponding points were not added to the figure due to the exact agreement.)

At the macroscale, the figure of merit *ZT* remains constant. Consequently, the maximum efficiency $\eta_{max}$ directly follows the variation of the temperature difference $\Delta T$ according to Eq.(4a)., increasing monotonically as *Transmissivity* decreases (Fig.3c). This suggests that lowering *Transmissivity* could geometrically enhance *TE* efficiency. However, this approach does not necessarily maximize overall *TE* performance due to the simultaneous increase in $R_{el}$, which can adversely affect the output power [Eq. (5)]. This trade-off is illustrated in Fig.4, which shows $P_{max}$ under three operating conditions: fixed $\Delta T$, experiment convective flow ($h$=200 Wm$^{-2}$K$^{-1}$), and weak convective flow ($h$=20 Wm$^{-2}$K$^{-1}$). In general, $P_{max}$ decreases with decreasing *Transmissivity* —unless the convective flow is weak, where it shows a non-monotonic variation. Notably, $P_{max}/A$ scales with the *Transmissivity* in all cases.

This behavior can be understood by considering Eq. (5), which shows that $P_{max}$ results from the interplay between the open-circuit voltage $V_{OC}$ and $R_{el}$. While $R_{el}$ increases with decreasing *Transmissivity* and is independent of operating conditions, $V_{OC}$ depends on $\Delta T$ (as $V_{OC} = S \cdot \Delta T$) (Fig.3d). Under fixed $\Delta T$, $V_{OC}$ is constant, so $P_{max}$ decreases with decreasing *Transmissivity* due to the increase in $R_{el}$. With decreasing $h$ (weaker convection), $\Delta T$ also drops, further reducing $P_{max}$. However, this reduction is less pronounced at lower *Transmissivity*, since $\Delta T$ is better preserved under weak convection when *Transmissivity* is low. As a result, although $P_{max}$ decreases under weak convective conditions compared to the fixed $\Delta T$ case, it can still remain higher than that of a *cuboid* leg. This reveals that a leg with *constricted* geometry may exhibit better TE performance in environments with weak convective heat transfer—but not under strong convection, where its higher resistance leads to reduced performance.

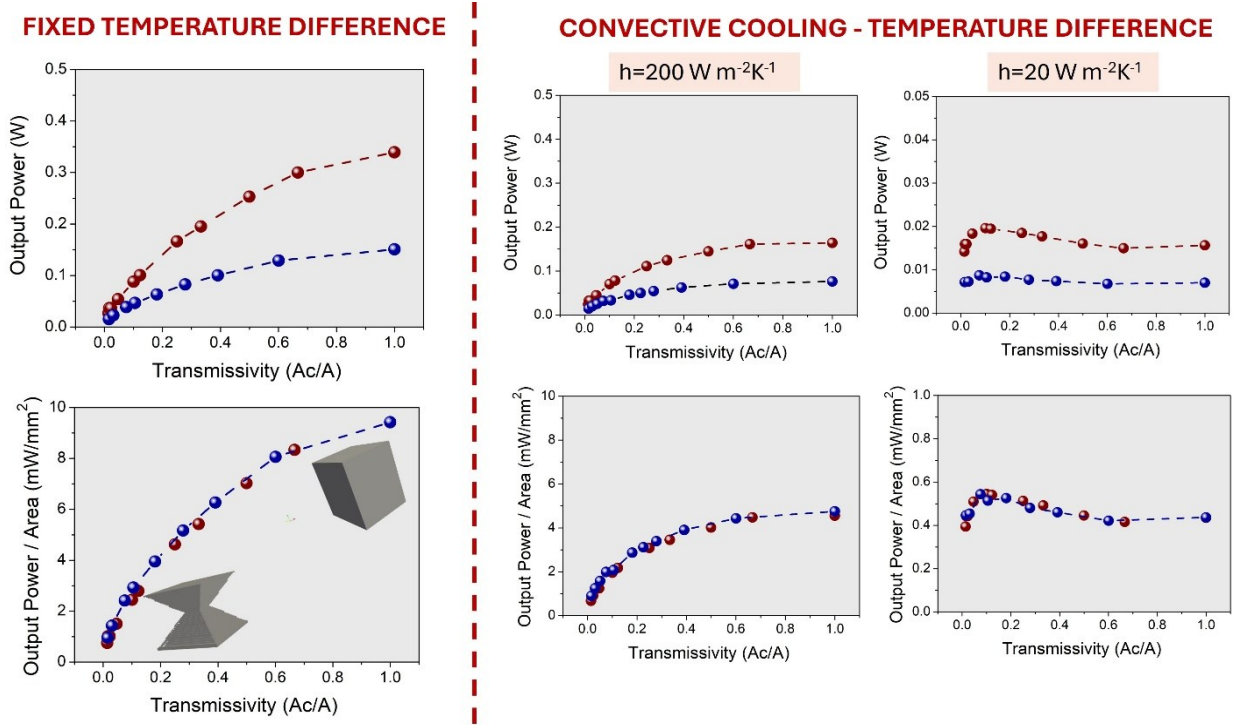

**FIG. 4. Output Power versus Transmissivity.** Maximum output power ($P_{max}$) and output power per area A ($P_{max}/A$) for the legs of Fig.3, under three operating conditions: fixed temperature difference (left panels), convective cooling with h=200 Wm$^{-2}$K$^{-1}$ (upper right) and weaker convective cooling (lower right panel) with h=20 Wm$^{-2}$K$^{-1}$.

## C. Universal performance relations with Transmissivity

Based on the previous analysis, universal relations of the electrical and thermal resistances, as well as key TE performance metrics, are derived in terms of *Tr*:

*Electrical and thermal resistances $R_{el(th)}$*

$$R_{el(th)} = R_o^{el(th)} F(Tr) \tag{21}$$

where *F(Tr)* is the resistance ratio $R/R_o$ of the *constricted* geometry to its uniform counterpart, applicable to both electrical and thermal transport. This function is independent of the type of current carriers (electrons or phonons), intrinsic material transport parameters and scattering mechanisms, and consequently material choice.

Eq, (21) was previously demonstrated for single and multiple sharp-constriction geometry profiles[32] and is here also confirmed for the hourglass geometry, the smooth-constriction case. $F(Tr)$ is plotted in Fig.3a for the hourglass geometry. These studies, in various *constricted* geometries, indicate that the functional dependence of $F(Tr)$ depends on the specific modulation profile. In *constricted* geometries, this functional exhibits the following scaling relation with $Tr$:

$$F(Tr) \sim Tr^{-n} \tag{22}$$

where $n \approx 1$ for multiple-constriction geometry modulation[32] and $n \approx 0.5$ for single-constriction geometry modulation, as shown in Fig.3a for the hourglass geometry.

### *Temperature difference $\Delta T$*

From Eqs. (20) and (21), it is obtained:

$$\Delta T\ (Tr) = \frac{hL/k}{F(Tr)^{-1} + hL/k} \Delta T_{max} \tag{23}$$

### *Maximum efficiency $\eta_{max}$*

From Eqs. (4a) and (21), it is obtained:

$$\eta_{max} = \frac{\Delta T_{max}}{T_h} \frac{\sqrt{1+ZT}-1}{\sqrt{1+ZT}+1-\frac{\Delta T_{max}}{T_h}\frac{hL/k}{F(Tr)^{-1}+hL/k}} \frac{hL/k}{F(Tr)^{-1}+hL/k} \tag{24}$$

### *Maximum output power $P_{max}$*

From Eqs. (5), (21) and (23) it is obtained:

$$P_{max} = P_{max}^o \left[\frac{hL/k}{F(Tr)^{-1} + hL/k}\right]^2 F(Tr)^{-1} \qquad (25)$$

where $P_{max}^o$ is the maximum output power of the corresponding cuboid leg under fixed temperature operating conditions ($\Delta T = \Delta T_{max}$):

$$P_{max}^o = \frac{S^2 \Delta T_{max}^2}{4 R_o^{el}} \qquad (26)$$

Eqs. (21)–(26) establish universal expressions for electrical and thermal resistances, as well as TE performance metrics, all formulated in terms of *Transmissivity*. These results demonstrate that every TE metric can be expressed in terms of a single scaling function of *Tr*, *F(Tr)*, representing the ratio $R/R_o$ of the constricted geometry to its uniform counterpart- independent of material properties or operating conditions. While *F(Tr)* depends on the specific constriction profile, all such geometries obey the scaling relations in Eq. (22), confirming that *Transmissivity* provides a unifying framework for performance comparison across diverse geometrical configurations.

**D. Unified optimization framework explaining observed output power enhancement**

The measured output power of composite legs was higher for the hourglass geometry compared to the cuboid leg of the same volume under the experimental conditions of fixed temperature difference and forced convection with $h$=200 W m$^{-2}$K$^{-1}$ [48]. This enhancement was initially attributed to the larger *GP*s parameters of the hourglass shape. However, calculations in Fig. 4 show a decrease in output power for the actual hourglass with *Tr*=0.086, as well as for any hourglass with Tr < 1, compared to the cuboid geometry (*Tr* = 1) under the same conditions. Thus, geometry modulation alone cannot explain the observed enhancement. To investigate this, we analyzed the experimental data in conjunction with calculations and derived analytical formalism. The results indicate that the dominant mechanism behind the observed power increase is the high contact resistance $R_c$ of the experimental composite legs.

The experimental output voltage and power versus current for composite legs are plotted in Fig.5. Data for the corresponding bare legs were obtained by calculating the bare leg's resistance $R_{b-leg}$ from the corresponding composite leg resistance $R_{c-leg}$ by subtracting the two contacts resistance $2R_c$, according to the equation:

$$R_{c-leg} = R_{b-leg} + 2R_c \qquad (27)$$

Conversely, the theoretical composite leg resistance $R_{c-leg}$ was obtained by adding the two contact resistances to the calculated bare leg resistance $R_{b-leg}$. Resistances $R$ can be extracted from the slope of the I-V curve, since $V = V_{OC} - IR$. The output power is given by $P = IV$. This allows direct comparison of all experimental and theoretical resistance, output voltage and power data for both composite and bare legs. The comparison in Fig.5 shows excellent agreement between calculations and measurements validating simulations.

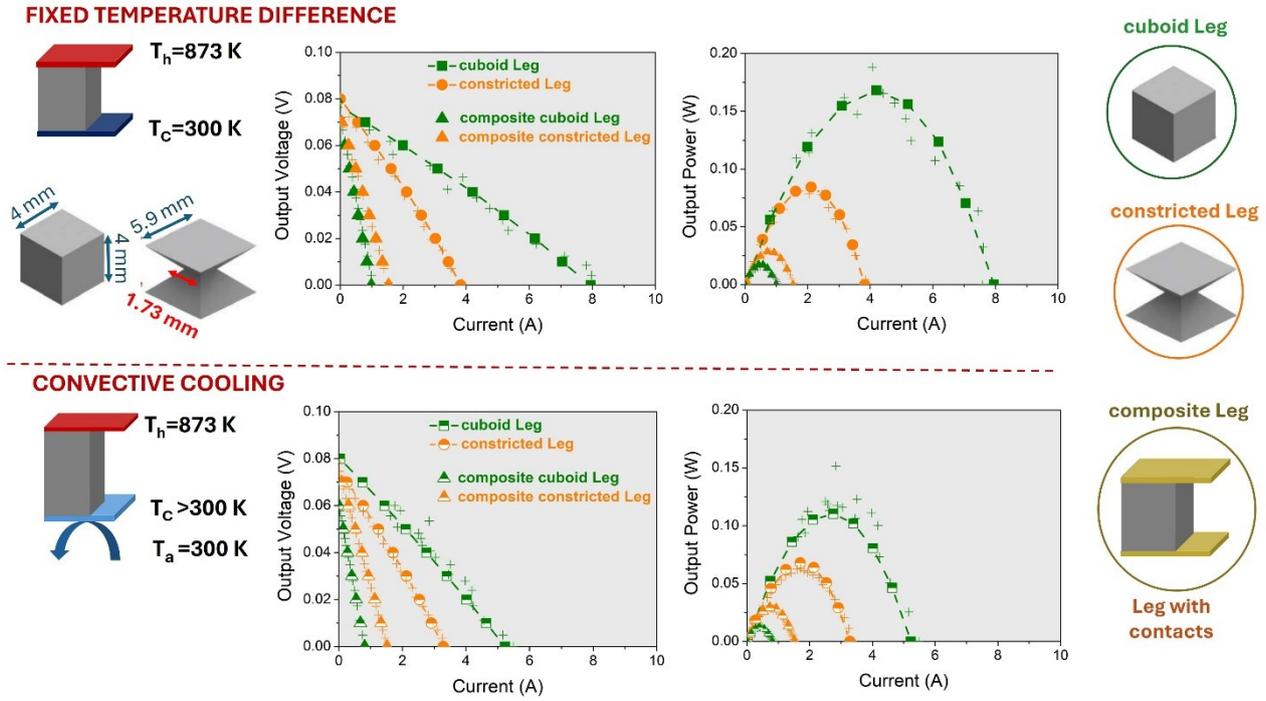

**FIG. 5. Output voltage and power** versus current for (bare) legs and composite legs under the experiment operating conditions: fixed temperature difference $\Delta T$ (upper panel) and forced convective cooling with $h$=200 Wm$^{-2}$K$^{-1}$ (lower panel). Calculations are shown with symbols. Experimental[48] data for composite legs and are shown with crosses. The corresponding data for bare legs, estimated by subtracting contact resistances, are also shown with crosses. $T_h$, $T_c$ and $T_a$ denote the temperature at the hot side, the cold side and ambient respectively.

Fig. 5 shows that (i) the output power of the hourglass bare leg is smaller than that of the bare cuboid leg, and (ii) the output power of the hourglass composite leg is higher than that of the composite cuboid leg at all current values under the experimental operating conditions. It is also evident that the output voltage and power of the composite legs deteriorate due to high contact resistance. This deterioration is quantified in the Table, which compares resistances and output powers of bare and composite legs in both absolute and relative terms under the experimental conditions. The maximum power $P_{max}$ of the cuboid (hourglass) composite leg is 91% (65%) lower than that of the cuboid bare leg at fixed $\Delta T$ and 89% (59%) lower under convective cooling with $h$=200 W m$^{-2}$K$^{-1}$. The decrease is smaller for the hourglass composite leg, which has a larger area $A$, smaller $Rc$, and consequently lower total resistance. The total contact resistance $R_{contacts}$ (from both contacts) constitutes 90% of the cuboid composite leg's electrical resistance, but only 60% of the hourglass composite leg's resistance. Therefore, the observed higher output power of the composite hourglass leg compared to the cuboid leg should not be interpreted as an enhancement due to geometry modulation, since such an enhancement would not occur if the contact resistance were lower. In contrast, calculations (Fig.6) for weaker convective cooling ($h=20$ W m$^{-2}$K$^{-1}$) show that the hourglass geometry outperforms the cuboid, and in this case, the enhancement can be attributed to the *constricted* geometry of the leg.

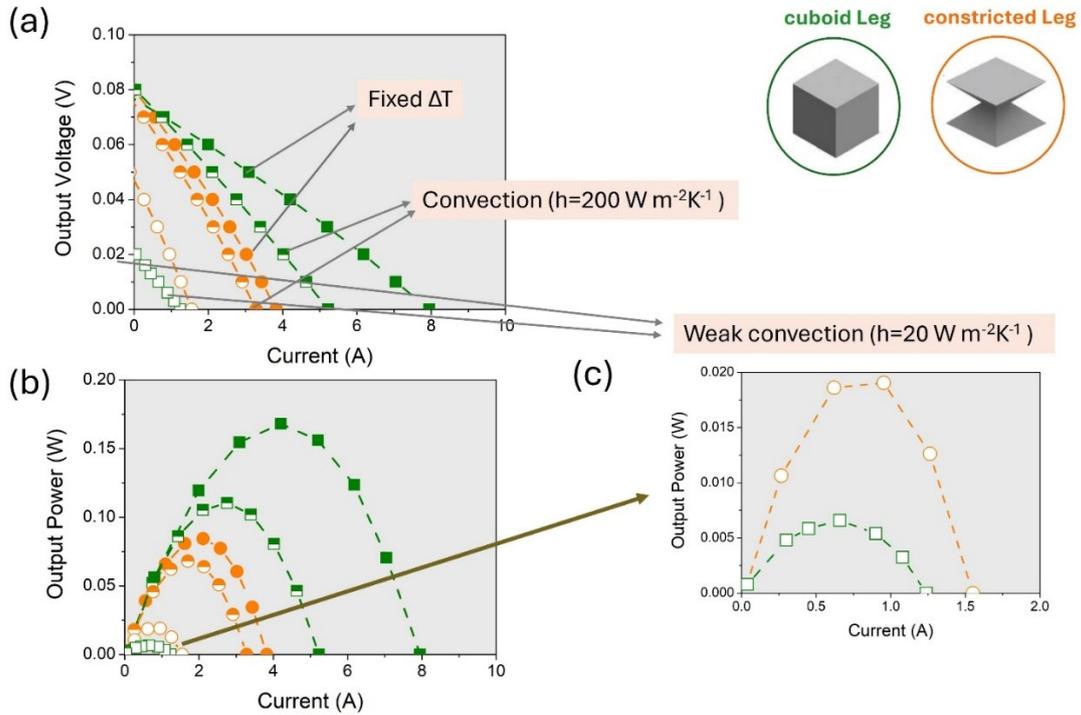

| LEG GEOMETRY | RESISTANCE | | Fixed ΔT | | | Convective cooling h=200 W m$^{-2}$ K$^{-1}$ | | |
|---|---|---|---|---|---|---|---|---|
| | | | Bare leg | Composite leg | BARE TO COMPOSITE | Bare leg | Composite leg | BARE TO COMPOSITE |
| | $R_{leg}$ (mΩ) | $R_{contacts}$ (mΩ) | $P_{max}$ (mW) | $P_{max}$ (mW) | | $P_{max}$ (mW) | $P_{max}$ (mW) | |
| cuboid | 7 (10%) | 61 (90%) | 168 | 15 | DECREASE 91% | 110 | 12 | DECREASE 89% |
| constricted (hourglass) | 19 (40%) | 28 (60%) | 84 | 29 | DECREASE 65% | 68 | 28 | DECREASE 59% |
| CUBOID TO CONSTRICTED | | | DECREASE 50% | INCREASE 93% | $P_{max}$ | DECREASE 38% | INCREASE 133% | $P_{max}$ |
| MAIN EFFECT | | | GEOMETRY MODULATION | CONTACT RESISTANCE | | GEOMETRY MODULATION | CONTACT RESISTANCE | |

**TABLE. Comparative Analysis of contact resistance and geometry modulation**. Resistance values of bare legs ($R_{leg}$) and contact resistances ($R_{contacts}$) are listed both in absolute terms and as percentages for the optimized experimental[48] structures. Corresponding output powers for bare and composite legs are shown under the experimental operating conditions. Bare leg performance is primarily influenced by geometry modulation, while composite leg performance is dominated by high contact resistance.

**FIG. 6. Operating conditions and power enhancement.** Calculated output voltage and power versus current for the bare legs shown in Fig.5 under three operating conditions: fixed temperature, convective cooling with $h=200$ Wm$^{-2}$K$^{-1}$, and weaker convective cooling with $h=20$ Wm$^{-2}$K$^{-1}$.

These findings underscore the importance of quantitatively characterizing the interaction between bare leg resistance and contact resistance in producing a performance maximum under varying operating conditions. Maximizing TE performance requires optimizing the output power of the device leg, which is governed by the composite leg resistance rather than solely by the saturation of $\Delta T$ (Eq. (5)). From Eqs. (5) and (27), it is obtained:

$$P_{max} = \frac{P_{max}^b}{1 + 2R_c/R_{b-leg}} \quad (28)$$

$$P_{max}^b = \frac{S^2 \Delta T^2}{4\, R_{b-leg}} \quad (29)$$

where $P_{max}$ and $P_{max}^b$ denote the maximum output power of the composite and the bare leg respectively under the specific operating conditions. Eq. (28) is an analytical relation of $P_{max}$ as a function of the ratio $R_c/R_{b-leg}$, providing insight to the interplay between contact resistance and the bare leg's resistance in determining the composite leg's maximum output power $P_{max}$. This relation serves as an analytical tool for examining how $R_{b-leg}$ and $R_c$ interact to produce maximum performance under any operating conditions. Fig.7 applies this analysis to the optimized hourglass geometry of the experiment.

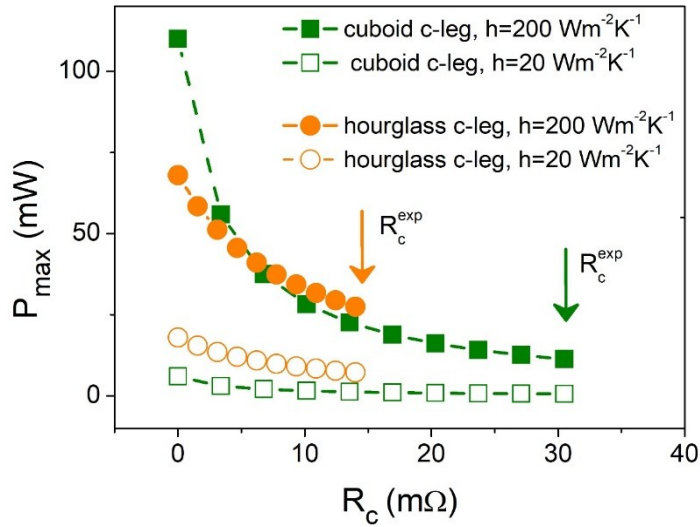

**Fig.7 Impact of contact resistance on TE performance.** $P_{max}$ versus contact resistance $R_c$, calculated using Eq. (28) for the experimental[48] optimized hourglass (orange) and cuboid (green) geometries, for $h$=200 Wm$^{-2}$K$^{-1}$ and $h$=20 Wm$^{-2}$K$^{-1}$. Experimental values of the contact resistance $R_c^{exp}$ are indicated by colored arrows for the cuboid (green) and hourglass (orange) geometries.

The results in Fig.7, show that $P_{max}$ decreases monotonically with increasing $R_c$. The value $P_{max}^b$ for the bare leg corresponds to $R_c=0$. For strong convection ($h=200$ Wm$^{-2}$K$^{-1}$), $P_{max}^b$ is higher for the cuboid geometry than for the hourglass geometry, whereas for weaker convection ($h=20$ Wm$^{-2}$K$^{-1}$), the hourglass geometry exhibits higher $P_{max}^b$. When $R_c \neq 0$, the influence of contact resistance is superimposed to the effect of geometry modulation, becoming progressively more significant and eventually dominant. For $h=200$ Wm$^{-2}$K$^{-1}$ the hourglass surpasses the cuboid for $R_c > 6$ mΩ, while for $h=20$ Wm$^{-2}$K$^{-1}$, the hourglass consistently outperforms the cuboid.

Separating the bare leg from the composite leg enables a concise, unified, area-dependent framework for practical device analysis. This involves two stages: (i) The bare leg is studied first to determine the constricted material resistances $R_{el}$ and $R_{th}$ thereby isolating the influence of *constricted* geometry on TE performance, free from device-scale effects such as the cross-sectional area dependence of $R_c$. (ii) The obtained $R_{el}$ can then be combined to $R_c$ through Eqs (28)-(29) to evaluate realistic device performance.

A unified optimization framework for the coupled system—comprising the contact electrodes and *constricted* material slab—is enabled by the identified scaling behavior with *Transmissivity*. In particular, the optimal $Tr_{opt}$ that maximizes the TE performance of the *constricted* material is directly obtained from the scaling plots. The optimal cross-sectional area $A_{opt}$ can be determined using standard optimization procedures for $R_c$ and $R_o$ of typical cuboid legs. Then, the optimal constriction area $A_C^{opt}$ follows from Eq.(1) as the product of these optimal values: $A_C^{opt} = A_{opt} Tr_{opt}$.

**E. Interpretation of geometry optimization**

As shown above, the geometry optimization of the hourglass leg can be effectively guided by the *Transmissivity*. The improved TE performance stems from the increased thermal resistance $R_{th}$, relative to that of the cuboid leg $R_o^{th}$. According to Eq. (20), this increase in thermal resistance leads to a higher temperature difference *ΔT* under convective operating conditions. Fig.3b illustrates that *ΔT* increases as *Tr* decreases. Furthermore, Fig.8 reveals a distinct trend when plotting *ΔT* against the *Inverse Transmissivity*: an initial sharp rise in *ΔT*, followed by a plateau that asymptotically approaches the ambient temperature as *Tr* becomes sufficiently low and $R_{th}$ becomes sufficiently high. Eq. (20) shows that the plateau height is determined by the convection coefficient *h*, the intrinsic thermal conductivity of the material *k*, and the leg length *L*. It also shows that the way *ΔT* levels off (as a plateau) with decreasing *Tr* is governed by the ratio $R_{th}/R_o^{th}$, i.e. the modulated geometry *Transmissivity*. Thus, the occurrence of the plateau is fundamentally a geometric effect arising from the *constricted* geometry.

Further analysis of the temperature profile (T-profile) across the leg provides insight into the physical mechanism behind the abrupt increase in thermal resistance $R_{th}$ as *Tr* $^{-1}$ (*Tr*) increases (decreases) within the plateau regime. Figs. 8a and 8b show that, at high *Tr*, the T-profile remains nearly linear. However, as *Tr* decreases, the profile begins to deform significantly and eventually develops a wide, stable temperature plateau centered at the constriction. According to Fourier's law, thermal conductivity is inversely related to the spatial temperature gradient. Therefore, the steep gradient observed at the constriction at the onset of the plateau indicates the emergence of a high thermal resistance that dominates $R_{th}$ of the leg. This resistance is the *Constriction Thermal Resistance (CTR)*, previously identified in width-modulated slabs with sharp constrictions[32]. Notably, the *CTR* remains nearly constant within the plateau, as evidenced by the invariant T-profiles across different values of *Tr* in this regime. This confirms that the formation of the *CTR* is responsible for the sharp increase in $R_{th}$ below a critical *Transmissivity*. Consequently, optimal TE performance is achieved at the onset of this plateau—marking the transition into the *Constriction Resistance Regime*—where both $R_{th}$ and $\Delta T$ reach favorable values.

In addition, Fig. 8c shows that the slope of the T-profile at the constriction remains unchanged when the constriction is shifted along the leg, indicating that *CTR* is invariant with respect to its vertical position. This is because *CTR* is a geometrical resistance determined by the constriction itself regardless of its position in the *constricted* leg. Since *CTR* dominates the overall thermal resistance $R_{th}$, legs with the constriction placed at different heights exhibit similar $R_{th}$ values and, consequently, comparable TE performance. This explains the previously unexplained observation that the output power of the optimized hourglass geometry depends only weakly on the vertical position of the constriction[48]—a result now attributed to the dominant role of *CTR*.

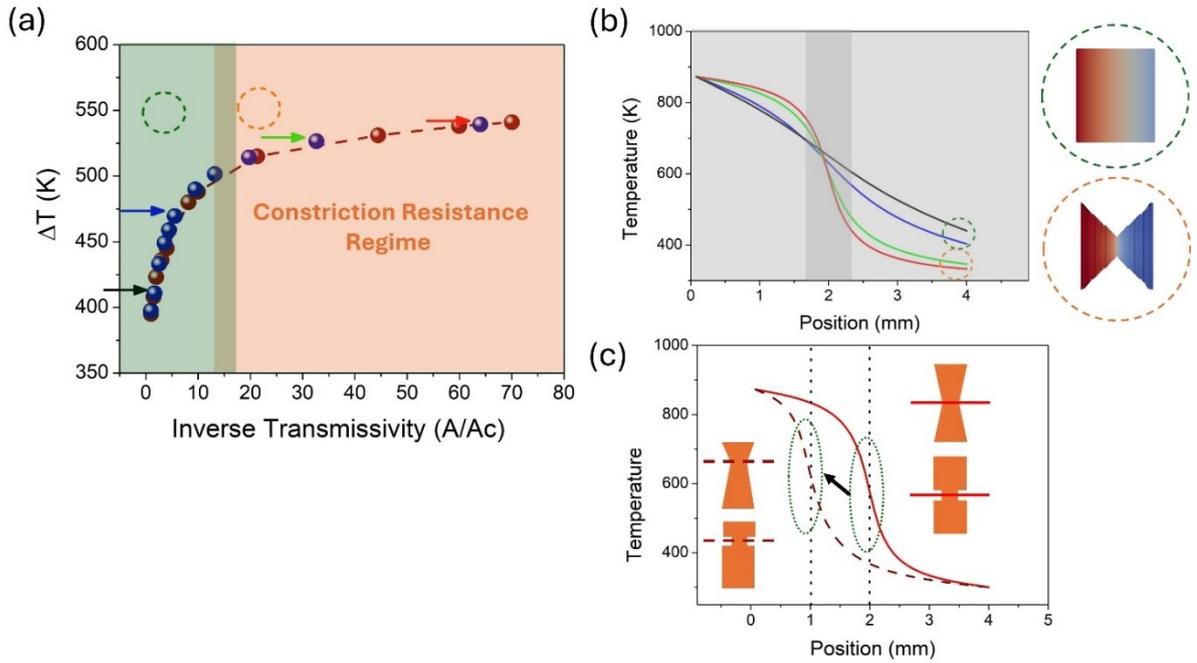

**FIG.8. Interpretation of geometry optimization.** (a) $\Delta T$ versus *Inverse Transmissivity* for h=200 Wm$^{-2}$K$^{-1}$; (b) T-profiles along the central axis of the leg corresponding to *Tr* values indicated by the colored arrows in (a); (c) T-profiles for the same *Tr* (red arrow) for two constriction positions: at the center (solid line) and the first quarter of the leg (dashed line).

## D. CONCLUSIONS

Thermoelectric metamaterials with widths modulated by constrictions have emerged as a promising design strategy for improving heat management and enhancing TE performance. Their potential has been experimentally validated across various non-cuboid TE leg geometries, where they have demonstrated superior performance. Calculations and analytical formalism show that this enhanced TE performance can be interpreted in terms of two mechanisms of geometric origin: (i) the *scaling behavior* of thermal resistance with *Transmissivity*, and (ii) the critical formation of the *Constriction Thermal Resistance CTR*. *Transmissivity* emerges as an intrinsic transport property of *constricted* geometries—rather than a simple geometric ratio.

A universal scaling formalism is derived for electrical and thermal resistance, as well as TE performance metrics, expressing these metrics in terms of a single scaling function of *Tr*, *F(Tr)*, which represents the relative resistance of the constricted geometry to the corresponding uniform one-independent of material choice or device operating conditions. While $F(Tr)$ depends on the specific constriction profile, all such geometries obey the scaling relation for the resistance ratio $R/R_o$ of the constricted geometry to its uniform counterpart, supporting the view that *Transmissivity* provides a unifying paradigm for comparing performance across distinct geometrical configurations.

It is demonstrated that the superior TE performance of the optimal hourglass geometry arises from its high thermal resistance dominated by the *CTR*. Moreover, the *GP* used to qualitatively interpret the experimental results is shown to be an inappropriate metric for guiding constricted leg design, as it does not capture the full effect of geometry modulation on electrical and thermal resistances. In contrast, this effect is entirely governed by *Tr* through its scaling relationship with transport properties, establishing *Transmissivity* as a robust descriptor of the impact of geometry modulation on TE performance.

Furthermore, a unified optimization framework is developed for composite TE legs, accounting for both the constricted material and the contact electrodes. This model reveals that previously observed power enhancements may largely stem from high contact resistance, rather than from *constricted* geometry alone. These findings underscore the importance of jointly optimizing both geometric design and contact fabrication to fully realize the performance potential of thermoelectric metamaterials.

Altogether, the findings of this work establish general optimization criteria and geometry-based performance guidelines for *constricted* TE legs, charting a promising pathway toward enhanced power generation through informed geometric engineering.

## Conflict of interest
The author has no conflicts to disclose.

## Data availability statement
The data that support the findings of this study are available within the article.

# Funding

This research received no external funding.